\documentclass[
nofootinbib,
amsmath,amssymb,
aps,
prd,
10pt
]{revtex4-2}

\usepackage[caption=false]{subfig}
\usepackage{graphicx}
\usepackage{dcolumn}
\usepackage{bm}
\usepackage{tcolorbox}
\usepackage[mathlines]{lineno}
\usepackage{ulem}
\usepackage{booktabs,multirow,tabularx,makecell}
\usepackage{enumitem}
\usepackage{todonotes}
\usepackage{slashed}
\usepackage{amsmath}
\usepackage{placeins}
\usepackage{booktabs,tabularx,ragged2e,enumitem,mathtools,bm}
\usepackage[percent]{overpic}
\newcolumntype{Y}{>{\RaggedRight\arraybackslash}X}

\setlist[enumerate]{leftmargin=1.2em,itemsep=2pt,topsep=2pt,parsep=0pt,partopsep=0pt}
\allowdisplaybreaks
\newcommand{\na}{\notag\\}

\newcommand{\dhd}{{\textstyle d} \lower.03ex\hbox{\kern-0.38em$^{\scriptstyle-}$}\kern-0.05em{}}
\usepackage{xcolor}

\usepackage[hidelinks]{hyperref}
\usepackage[commandnameprefix=ifneeded]{changes}

\begin{document}
\title{Comparative Study of Color Glass Condensate and Collinear Frameworks for Large-Transverse-Momentum Semi-Inclusive Deep Inelastic Scattering}
%
%
\author{Swagato Mukherjee}
\affiliation{Physics Department, Brookhaven National Laboratory, Upton, New York 11973, USA}
\author{Bj\"orn Schenke}
\affiliation{Physics Department, Brookhaven National Laboratory, Upton, New York 11973, USA}
\author{Shaswat Tiwari}
\email{stiwari1@bnl.gov}
\affiliation{Physics Department, Brookhaven National Laboratory, Upton, New York 11973, USA}
\begin{abstract}
We study the Semi-Inclusive Deep Inelastic Scattering (SIDIS) cross-section at large values of hadron transverse momentum $P_t \gtrsim Q$, with $Q^2$ being the photon virtuality. This kinematic regime of SIDIS allows for a collinear factorization in terms of parton distribution functions (PDFs) and collinear fragmentation functions (FFs). On the other hand, at high energies, i.e., at small Bjorken $x$, the same SIDIS process can also be factorized within the Color Glass Condensate (CGC) framework in terms of eikonal dipole amplitudes. In this work, we perform a systematic comparison of the leading-order (LO) and next-to-leading-order (NLO) collinear and the LO CGC factorizations of this process based on the COMPASS and HERA data. We further provide comparisons of these two factorization frameworks for nucleon and nuclear large-$P_t$ SIDIS in the expected Electron Ion Collider (EIC) kinematics. We find, within the present theoretical uncertainties, that it might be difficult to distinguish between these two factorization schemes for nuclear SIDIS at the EIC. We also observe a significant dependence of the LO eikonal CGC predictions on the choice of longitudinal momentum fraction $x_g$, suggesting significant beyond-eikonal corrections for the EIC kinematics.

\end{abstract}
%
\date{\today}
\maketitle
%
%
%
\section{Introduction}
Quantum Chromodynamics (QCD) factorization allows us to study the non-perturbative content of hadronic states involved in a scattering reaction.
It is formalized by factorization theorems \cite{Collins:1981uk,Collins:1985ue,Collins:1988ig,Collins:1989gx}, that allow one to write a process, at a given kinematics, in terms of perturbatively computable hard factors and non-perturbative operators. The hard-factors are process dependent while the non-perturbative operators are considered ``universal'', i.e., they appear in the factorization of different processes. However, these ``universal'' operators are intrinsically tied to the kinematic regime in which the corresponding factorization theorem applies. In a different kinematic regime, the same process may instead factorize in terms of a different set of ``universal'' operators.
We shall study two such descriptions, 1) collinear factorization \cite{Daleo:2004pn,Gonzalez-Hernandez:2018ipj,Kniehl:2004hf,Wang:2019bvb,Whitehill:2022mpq,Boglione:2019nwk} and 2) Color Glass Condensate (CGC) factorization \cite{Marquet:2009ca,Iancu:2020jch,Bergabo:2022zhe,Bergabo:2024ivx,Altinoluk:2024SIDIS,Radhakrishnan:2026bxy} of Semi-Inclusive-Deep-Inelastic-Scattering (SIDIS), where both the longitudinal and transverse momenta of one outgoing hadron are measured.

It has been shown that SIDIS admits a collinear factorization \cite{Daleo:2004pn,Gonzalez-Hernandez:2018ipj,Kniehl:2004hf,Wang:2019bvb,Whitehill:2022mpq,Boglione:2019nwk} in terms of parton distribution functions (PDFs) when the transverse momenta of the outgoing hadron $P_t \gtrsim Q$, where $Q^2$ is the virtuality of the photon and the hard scale of the process. The hadronization process is captured by the so-called 
collinear
fragmentation functions (FFs). This factorization in terms of PDF and FF has been proven to all orders in \cite{Ellis:1978ty,Collins:1989gx,Collins:1981uw} and the hard factor has been computed to next-to-leading order (NLO) \cite{Daleo:2004pn,Wang:2019bvb}. For an electron scattering on a proton, it has been shown that NLO SIDIS provides a 
reasonable description of the HERA H1 data \cite{Daleo:2004pn,Gonzalez-Hernandez:2018ipj}. Similarly, a study of the COMPASS measurement \cite{COMPASS:2017mvk} for scattering on deuterons showed that NLO SIDIS provided a reasonable description of the experimental data \cite{Gonzalez-Hernandez:2018ipj}. 

On the other hand, it is known that as we increase the center of mass energy at fixed $Q^2$ and approach the Regge limit, the higher twist PDF operators start playing a bigger role \cite{Weigert:2005us,Bartels:2000hv}. Thus the leading-twist collinear factorization is expected to break down in the Regge limit as we approach smaller values of Bjorken-$x$, $x_B$. This is the regime of the Color Glass Condensate (CGC) effective theory \cite{McLerran:1994CGC1,McLerran:1994CGC2,McLerran:1994CGC3,Balitsky:1996HighEnergy,Kovchegov:1999BK,JalilianMarian:1997BFKL,JalilianMarian:1998JIMWLK1,JalilianMarian:1998JIMWLK2,Iancu:2001CGC,Weigert:2002JIMWLK,Iancu:2003xm,Gelis:2010nm,Caucal:2025zkl}.

In the Regge limit, CGC effective theory can be used to factorize Deep Inelastic Scattering (DIS) cross sections in terms of dipole amplitudes. These amplitudes are all-twist operators constructed from infinite Wilson lines and constitute the leading operators in the eikonal expansion. This framework was instrumental in establishing geometric scaling in inclusive DIS, whereby the cross section depends on the photon virtuality and Bjorken-$x$ predominantly through a single scaling variable involving the saturation scale.
It leads to the prediction of gluon saturation \cite{McLerran:2001sr,Iancu:2003xm,Weigert:2005us,Kovner:2005ja,
JalilianMarian:2005jf,Gelis:2010nm,Albacete:2014fwa,Mueller:2026GluonSaturation}, which is one of the core objectives of the upcoming Electron-Ion-Collider (EIC)\cite{Accardi:2012qut}.
    
SIDIS has been computed up to NLO order in CGC effective theory \cite{Marquet:2009ca,Iancu:2020jch,Bergabo:2022zhe,Bergabo:2024ivx,Altinoluk:2024SIDIS,Duan:2026jrk}. The hadronization is done through collinear fragmentation functions. For phenomenological applications, we shall however restrict ourselves to LO SIDIS. The initial condition for the dipole amplitudes are computed using the McLerran-Venugopalan (MV) \cite{McLerran:1994CGC1,McLerran:1994CGC3,McLerran:1994CGC2} model with the evolution being done using the running-coupling Balitsky-Kovchegov (rcBK) \cite{Balitsky:2006wa,Kovchegov:2006vj,Albacete:2004gw,Albacete:2010sy,Casuga:2026dif} framework. Using this methodology we compute the SIDIS cross-section in the CGC and compare with the collinear predictions for COMPASS \cite{COMPASS:2017mvk} and HERA H1 \cite{H1:2004xgw} kinematics at data points with $x_B \leq 0.01$.  \par

We further compare the predictions of collinear and CGC factorizations for semi-inclusive scattering on nuclei. In collinear factorization we compute the SIDIS cross-section using the nuclear parton distribution functions (nPDFs) \cite{Kovarik:2015cma,Duwentaster:2022kpv,AbdulKhalek:2019mzd,
AbdulKhalek:2020yuc,Eskola:2021nhw,Walt:2019slu,Helenius:2021tof,
Hou:2019efy}, while in the CGC framework we use an $A^{1/3}$ scaling of the squared saturation scale with the mass number $A$.

The paper is organized as follows. We begin with a review of factorization of SIDIS in both collinear and CGC frameworks in Section \ref{sec:SIDIScross-sections}. This is followed by a comparison of the two cross-sections for COMPASS kinematics in Section \ref{sec:COMPASS} and for HERA kinematics in Section \ref{sec:HERA}. We then move on to an analysis of nuclear SIDIS in Section \ref{sec:nuclear}. We summarize our results in Section \ref{sec:conclusion}. Technical details are provided in the appendices.

\section{SIDIS cross-sections} \label{sec:SIDIScross-sections}
\begin{figure}[tbp]
    \centering
    \includegraphics[width=0.7\textwidth]{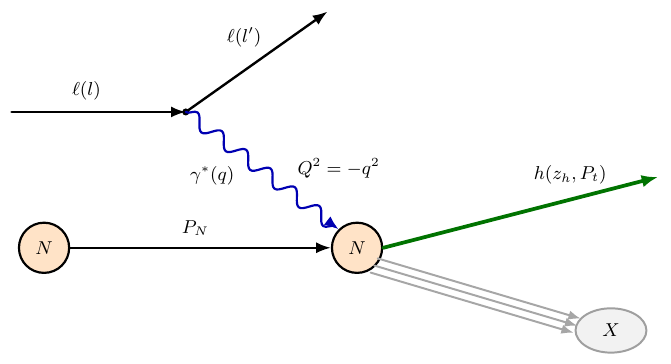}
    \caption{A representation of SIDIS, where a lepton with momentum $l$ scatters with a hadron/nucleus with momentum $P_N$. The momenta of the outgoing lepton and one hadron are measured. Undetected particles are represented by $X$.}
     \label{fig:SIDIS}
\end{figure}

In semi-inclusive deep inelastic scattering we scatter a lepton on a hadron/nucleus and measure the momentum of the outgoing lepton and hadron (see Fig.\,\ref{fig:SIDIS}). The differential SIDIS cross-section has the form
\begin{align}
\label{eq: cross_section}
\frac{d \sigma^{l N\:\rightarrow\:lhX}}{dQ^2 d x_B d z_h dP_t^2}
\end{align}
where $x_B$ is the Bjorken-$x$ variable, $Q^2$ is the photon virtuality, $z_h$ is the longitudinal momentum fraction of the hadron, $P_t$ is measured transverse momentum of the hadron and $N$ denotes the incoming nucleus/hadron. Note that we integrate over $\phi_h$, the angle made by the hadron with respect to the lepton plane. The cross-section is computed at fixed inelasticity $y$. Some useful relations between these kinematically relevant quantities can be found in Appendix \ref{app:Conv}. Note that we neglect the lepton mass, quark mass, target mass and the mass of the detected hadron in this analysis. 

We are interested in the SIDIS cross-section of Eq.\,\eqref{eq: cross_section} at large values of $P_t \gtrsim Q$, where the photon virtuality Q is the hard scale of the process. It is well known that this kinematic regime admits a collinear factorization in terms of PDF of the incoming hadron/nucleus and the FF of the outgoing hadron. Alternatively, at small $x_B$ or large center of mass energy, a complementary factorization scheme in terms of dipole amplitudes constructed from Wilson lines can be used. This is the Color Glass Condensate effective theory. We review each of these formalisms below before proceeding to compare the two factorization schemes.
\FloatBarrier
\subsection{Collinear factorization}
At moderate values of $x_B \gtrsim 0.1$, the typical transverse momentum carried by a parton is $\mathcal{O}(\Lambda_{\rm QCD})$. This implies that the large transverse momentum of the outgoing hadron $P_{t}\gtrsim Q$ comes from the hard quark-gluon splitting, in the frame where the virtual photon transverse momentum $q_t\,=\,0$. As a result, the first non-zero cross-section for large-$P_t$ SIDIS appears at $\mathcal{O}(\alpha_s)$. Furthermore, the large transverse momentum of the outgoing hadron is not sensitive to the intrinsic transverse momentum of the partons in the incoming hadron and the transverse momentum gained by the fragmenting hadron during the fragmentation process since both momenta are of $\mathcal{O}(\Lambda_{\rm QCD})$. This implies that the relevant non-perturbative objects are parton distribution functions (PDFs) and collinear fragmentation functions (FFs) as opposed to transverse momentum dependent distributions (TMDs) at $P_t \ll Q$ \cite{Boussarie:2023izj}. Higher order corrections correspond to more gluon/quark emissions. The general SIDIS cross-section (see \cite{Daleo:2004pn}) in collinear factorization at large $P_t$ is thus given by
\begin{align}
\label{eq:SIDIS_collinear}
\frac{d \sigma^{l N\:\rightarrow\:lhX}}{dQ^2 d x_B d z_h dP_t^2}\,=\,\sum_{abn}\:\frac{1}{z_h (Q^2 +W^2)}\:\int_{z_h}^{z_{\rm max}}\:\frac{dz}{1-z}\:f_{a}(\xi)\:D_{h/b}\left(\frac{z_h}{z}\right)\:\frac{d\sigma_{ab}^{n}}{dQ^2 dx_B dz} 
\end{align}
where
\begin{align}
\label{eq:y_max_and_xi}
z_{\rm max}\,=\frac{Q^2\,(1-x_B)\,z_h^2}{Q^2 (1-x_B) z_h^2+ P_t^2 x_B} \,, \qquad \xi = \frac{z_h^2 Q^2 (1-z) +  z P_t^2}{ z_h^2(Q^2 + W^2) (1-z)} \,,
\end{align}
and $a$ and $b$ denotes the parton scattering and fragmenting,  respectively. The index $n$ denotes the $\alpha_s$ order being considered. The kinematic constants are defined in Appendix \ref{app:Conv}. The variable $z$ can be understood as the fraction of longitudinal momentum of the photon that is carried by the fragmenting quark. The partonic scattering cross-section is given by
\begin{align}\label{eq:psigma}
&\frac{d\sigma^{1}_{qq}}{dx_B\, dQ^2\,dz}\,=\,\frac{4\pi \alpha_{em}^2 e_q^2}{\xi x_B^2 s^2}\,C_F\,\bigg[ \bigg(\frac{(x_B + \xi z)^2 + 2 \xi (\xi - x_B - \xi z)}{\xi (\xi - x_B)(1-z)}\bigg)\,Y_T\,+\,2\,\frac{x_B\,z}{\xi}\,Y_L \bigg]\,\na &
\frac{d\sigma^{1}_{qg}}{dx_B\, dQ^2\,dz}\,=\,\frac{4\pi \alpha_{em}^2 e_q^2}{\xi x_B^2 s^2}\,C_F\,\bigg[ \bigg(\frac{\xi^2 + (x_B - \xi z)^2}{\xi (\xi - x_B) z}\bigg)\,Y_T\,+\,2\frac{x_B\,(1-z)}{\xi}\,Y_L \bigg] \na &
\frac{d\sigma^{1}_{gq}}{dx_B\, dQ^2\,dz}\,=\,\frac{4\pi \alpha_{em}^2 e_q^2}{\xi x_B^2 s^2}\,T_F\,\bigg[ \bigg(\frac{\xi^2 - 2 x_B (\xi - x_B) - 2 \xi^2 z (1-z)}{\xi^2 z (1-z)}\bigg)\,Y_T\,+\,\frac{4 (\xi - x_B) x_B}{\xi^2}\,Y_L \bigg]
\end{align} 
where the kinematic polarization factors are given by
\begin{align}
Y_T\,=\,\frac{1+(1-y)^2}{2y^2}\,, \qquad Y_L\,=\,\frac{1 + 4 (1-y) + (1-y)^2}{2y^2} \,,
\end{align}
and the constants $C_F$ and $T_F$ are the Casimir and trace for the fundamental representation of SU($N$), respectively. We shall also work with the $\mathcal{O}(\alpha_s^2)$ cross-section, which was computed and numerically implemented in \cite{Daleo:2004pn, Gonzalez-Hernandez:2018ipj}. We do not explicitly write it for brevity. As discussed, collinear factorization assumes that there is no large transverse momentum transfer from the incoming hadron/nucleus. This assumption becomes contentious at small values of $x_B$, where multiple scattering becomes important. This regime is understood through the CGC effective theory. 

\subsection{Small-$x$ CGC factorization}
The CGC is an effective theory of large longitudinal momentum modes (valence modes) used to study electron-hadron (nucleus) and hadron (nucleus)-hadron (nucleus) scattering at high energies. The incoming hadron (target) state is defined through the weight functional $W[\rho]$, which provides the distribution of color charges $\rho$ inside the hadron. These color charges generate the soft fields that scatter with the incoming projectile (photon in SIDIS) to yield Wilson lines under the leading (eikonal) approximation for large center-of-mass energy. In the center of mass frame of the virtual photon-target scattering, the target appears Lorentz contracted. Thus, CGC calculations involve scattering of the projectile on a hadronic/nuclear shockwave. For SIDIS, such a procedure involves the initial splitting into quark-antiquark, determined by the virtual photon wave function, followed by scattering with the shockwave. At leading order the cross-section can be written in terms of dipole amplitudes
\begin{align}
\label{eq:dipole_amplitude}
S(x_g,\textbf{r},\textbf{R})\,=\,\frac{1}{N_c}\,{\rm Tr}(V^{\small \dagger}_{\textbf{x}_1} V^{}_{\textbf{x}_2})\bigg|_{x_g}\,,
\end{align}
where the Wilson line $V_{\textbf{x}}$ is given by
\begin{align}
V_{\textbf{x}}\,=\,\exp\left(ig\int_{-\infty}^{\infty}\,dz^-\,t^aA_-^a(z^-,\mathbf{x})\right)\,,
\end{align}
where $t^a$ is the fundamental generator of SU($N$) and $\mathbf{r} =\textbf{x}_1 - \textbf{x}_2  $ is the relative co-ordinate and $\mathbf{R} = \frac{1}{2} (\textbf{x}_1 + \textbf{x}_2)$ is the center of mass coordinate of the dipole.

To compute the SIDIS cross-section in the CGC framework, the appropriate point to start is the DIS dijet cross-section shown in \cite{Iancu:2020jch,Bergabo:2022zhe,Bergabo:2024ivx} and integrate out either quark or anti-quark. This procedure is detailed in Appendix \ref{app:dijet}. For the analysis presented in this work, we assume that the dipole
$S$-matrix is isotropic and independent of the impact parameter and write
\[
S(x_g,\mathbf{r},\mathbf{R}) \equiv S(x_g,r),
\]
where the transverse extent of the hadron or nucleus is encoded through
a finite area $A_N = \pi R_N^2$. We further identify the small-$x_g$
evolution variable with Bjorken $x_B$, $x_g = x_B$, following the
prescription adopted in Ref.~\cite{Lappi:2013zma}. To assess the
sensitivity of our results to this approximation, we also consider an
alternative, transverse-momentum-dependent choice
$x_g = x_g(P_t,Q_s)$, defined in Eq.~\ref{eq:P_t suppress}. The cross-section for SIDIS under this approximation reads
\begin{equation}
\label{eq:SIDIS_small_x}
\begin{aligned}
{d \sigma^{e N\:\rightarrow\:ehX} \over dQ^2\,dx_B\,dz_h\:dP_t^2}\:=\:\sum_{\lambda\:=\:L,T}\:\tilde{Y}_{\lambda}(x_B, Q^2)\:{d \sigma^{\gamma^*_\lambda N\:\rightarrow\:hX} \over d z_h\:d P_t^2}\,=\,\sum_{b}\int_{z_h}^1\,\frac{dz}{z_h^2}\,z\sum_{\lambda\,=\,L,T}\,\tilde{Y}_{\lambda}(x_B, Q^2)\,D_{h/b}(\frac{z_h}{z})\,{d \sigma^{\gamma^*_\lambda N\:\rightarrow\:bX} \over d z\:d k^2} \,,
\end{aligned}
\end{equation}
where the leptonic factors are given by
\begin{align}
\tilde{Y}_L\,=\,\frac{\alpha_{\rm em} (1-y)}{\pi x_B Q^2}\,,
\qquad
\tilde{Y}_T\,=\,\frac{\alpha_{em} (1+ (1-y)^2)}{2\pi x_B Q^2}\,,
\end{align}
and $k = \frac{z}{z_h} P_t$. The sum b runs over both quarks and antiquarks. The virtual photon cross-section can now be written as 
\begin{align}
\label{eq:SIDIS_partonic_CGC}
&{d \sigma^{\gamma_L^* N \rightarrow qX} \over d z\:dk^2}\:\na &=\:\frac{2 \epsilon_f^2N_c\alpha_{em}e_q^2\,\mathcal{A}_N\,z(1-z)}{\pi}\:\left(\frac{1}{(k_\perp^2 +\epsilon_f^2)^2}-\:\:\int_0^\infty r\:d r\:J_0(k r)\Big[2K_0(\epsilon_f r)\:\frac{1}{k_\perp^2 +\epsilon_f^2}\:-\frac{r}{2\epsilon_f}\:K_1(\epsilon_f r)\Big]S(x_g,r)\right)\,, \na&
{d \sigma^{\gamma_T^* N \rightarrow qX} \over d z\:dk^2}\:\na &=\:\frac{N_c\alpha_{em}e_q^2\,\mathcal{A}_N\,\big[z^{2} + (1-z)^2\big]}{2\pi}\:\Bigg(\frac{k^2}{(k^2 + \epsilon_f^2)^2} + \epsilon_f\,\int_0^{\infty} r dr\:\:\Big[-\frac{2\:k\:K_1(\epsilon_f r)\:J_1(kr)}{\: (k^2 + \epsilon_f^2)}\:+\:\frac{J_0(kr)}{\epsilon_f}\na &\bigg(K_0(\epsilon_f r)  - \frac{r\epsilon_f}{2} K_1(\epsilon_f r)\bigg)\Big]S(x_g,r)\Bigg)\,,
\end{align}
where $\epsilon_f^2\,=\,z(1-z)Q^2$ and $\mathcal{A}_N$ is the area of the nucleus. We use $\mathcal{A}_N\,=\,A^{2/3} \mathcal{A}_p$, where $\mathcal{A}_p$ is the area of the proton and $A$ is the mass-number. For numerics, we shall use the MV model expression \cite{McLerran:1994CGC1,McLerran:1994CGC3,McLerran:1994CGC2} for the dipole at initial $x_0$
\begin{align}
S(x_0,r)\,=\,\exp\left[-\frac{\left(r^2\,Q_{sN}^2\right)^{\gamma}}{4}\,\ln\left(\frac{1}{r \Lambda}+e_c\right)\right]\,.
\end{align}
We choose the parameters $\gamma = 1$, $e_c = 1$ and $\mathcal{A}_p\,=\,18.81\,mb$ obtained in \cite{Kowalski:2007rw, Lappi:2013zma} by fitting to fixed-target and HERA data \cite{H1:2009pze}, while varying $Q_{s0}^2 \in [0.104, 0.416]\,{\rm GeV^2}$. The nuclear saturation scale is taken to be $Q_{sN}^2\,=\,A^{1/3} Q_{s0}^2$. Initializing at $x_0 = 0.01$, we use running-coupling Balitsky-Kovchegov (RCBK) equation to evolve the dipole distribution up to the required $x_g$ with $C^2 = 14.5$ (see \cite{Kowalski:2007rw, Lappi:2013zma}).  We shall now compare the cross-section for the two factorization schemes  Eq.\,\eqref{eq:SIDIS_small_x} and Eq.\,\eqref{eq:SIDIS_collinear} in COMPASS and HERA kinematics.

\section{COMPASS kinematics}\label{sec:COMPASS}
In this section we compare the measurements made by the COMPASS Collaboration  \cite{COMPASS:2017mvk} with the computed cross-sections introduced in the previous section. COMPASS measured the charged hadron multiplicities produced in a muon-deuteron scattering experiment. The multiplicities are defined as
\begin{align}
\label{eq:multiplicity}
\frac{d^2M^h(Q^2,x_B,z_h,P_t^2)}{dz_h\,dP_t^2}\,=\,{d \sigma \over dQ^2\,dx_B\,dz_h\:dP_t^2}\bigg/\frac{d\sigma^{\rm dis}}{dQ^2\,dx_B}\,,
\end{align}
where both SIDIS and DIS cross-sections are evaluated in the same factorization scheme. The DIS cross-section for the collinear factorization framework was evaluated using the package \cite{Candido2024Yadism}. The DIS cross-section in the CGC framework was evaluated following \cite{Lappi:2013zma}. For collinear SIDIS, the average of the proton and neutron PDF was used for the deuteron PDF:
\begin{align}
f_{d}(\xi)\,\approx\,\frac{1}{2}(f_p(\xi) + f_n(\xi) )\,,
\end{align}
where the PDF is taken from CT18 PDFs \cite{Hou:2019efy}. The fragmentation function is taken from \cite{Bertone:2017tyb}. Since LO collinear SIDIS is $\mathcal{O}(\alpha_s)$, we use NLO PDF and NLO FF in the cross-section, while for NLO collinear SIDIS we use NNLO PDF and NNLO FF in the computation.
\begin{figure}[tbp]
    \centering
    \includegraphics[width=\textwidth]{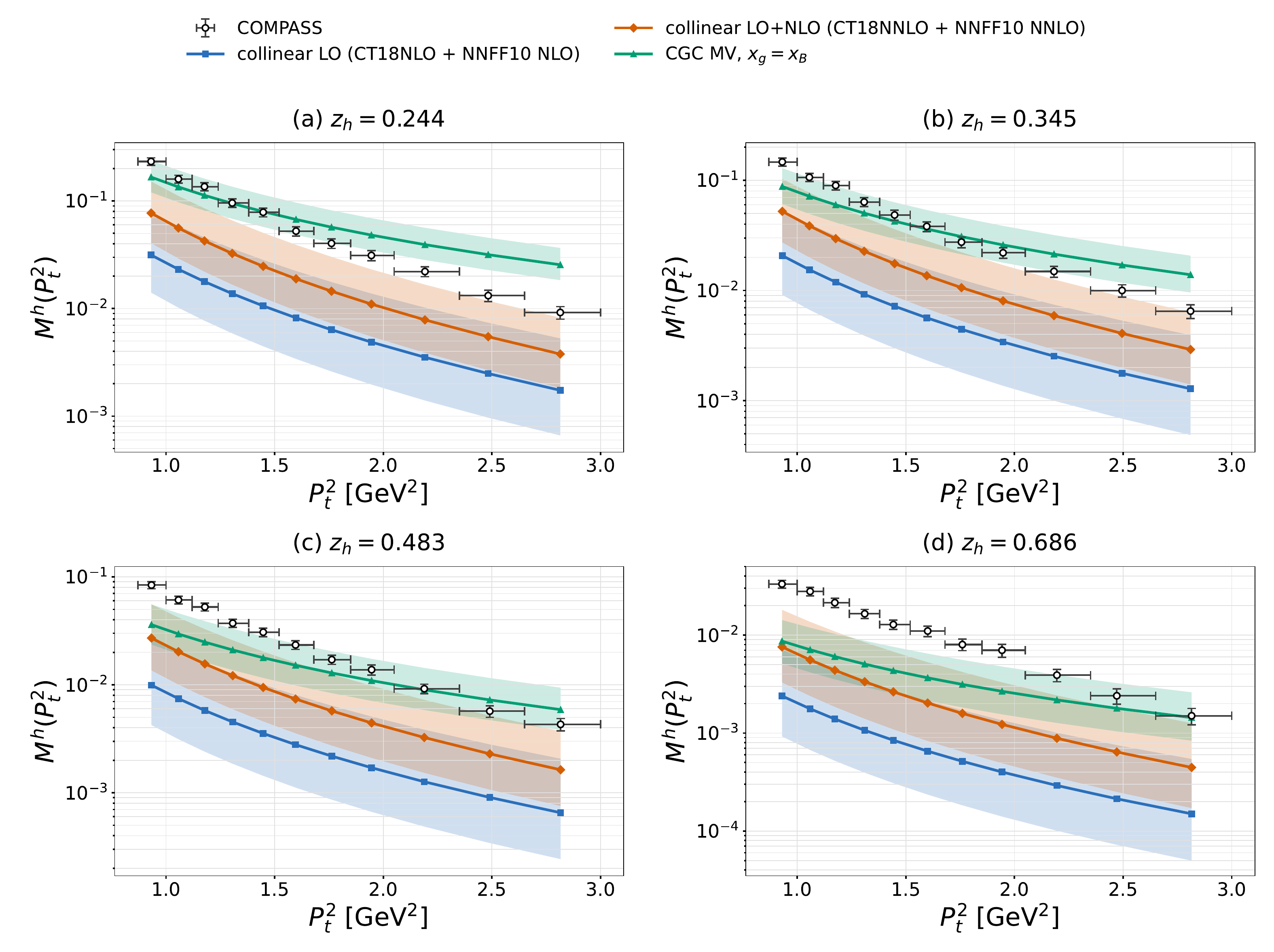}
    \caption{Comparisons to COMPASS data at $Q^2=1.3\,{\rm GeV}^2,\,x_B\,=\,0.0062$ and $y=0.69$ (computed at central values without bin averaging) \cite{COMPASS:2017mvk}. We use the CT18NLO/CT18NNLO \cite{Hou:2019efy} fit as our PDF and NNFF10NLO/NNFF10NNLO \cite{Bertone:2017tyb} as our fit for the FF in LO and NLO collinear cross-sections, respectively. Uncertainty bands for collinear results are obtained by combining PDF,FF replica uncertainty and  scale variation $\mu = r Q$ with $r \in [1.135,4]$. The CGC uncertainty band is constructed from a combination of varying $Q_{s0}^2 \in [0.104,0.416]\,{\rm GeV}^2$ and FF scale variation by factors of two around the central value. The bold lines correspond to the choices $\mu = 2Q$ and $Q_{s0}^2 = 0.208 {\rm GeV}^2$. }
    \label{fig:multiplicity_COMPASS_without_ymax}
\end{figure}

In the CGC calculation, we use the MV model \cite{McLerran:1994CGC1,McLerran:1994CGC2,McLerran:1994CGC3} initial condition with nuclear mass number $A=2$. The fragmentation function is taken at NLO and is the same as in the collinear cross-section \cite{Bertone:2017tyb}. 

The error bands are constructed by propagating all PDF replicas, either Monte-Carlo or Hessian eigenvector sets (available in LHAPDF \cite{Buckley:2014ana}) through the cross-section calculation. At each choice of factorization scale, the 90 percent confidence interval is calculated. Additionally, we also vary the factorization scale $\mu_F = rQ$ with $r \in [1.135,4]$. Note that the choice of $r_{\rm min} = 1.135$ is dictated by the minimum $\mu_F$ for CT18 PDFs. Specifically for the CGC curves we further vary $Q_{s0}^2 \in [0.104,0.416]\,{\rm GeV}^2$ to construct the error bands.  Choosing the smallest available value of $x_B = 0.0062$, we present the resulting comparison plot for the multiplicity in Fig.\,\ref{fig:multiplicity_COMPASS_without_ymax}. 

\begin{figure}[tbp]
    \centering
    \includegraphics[width=\linewidth]{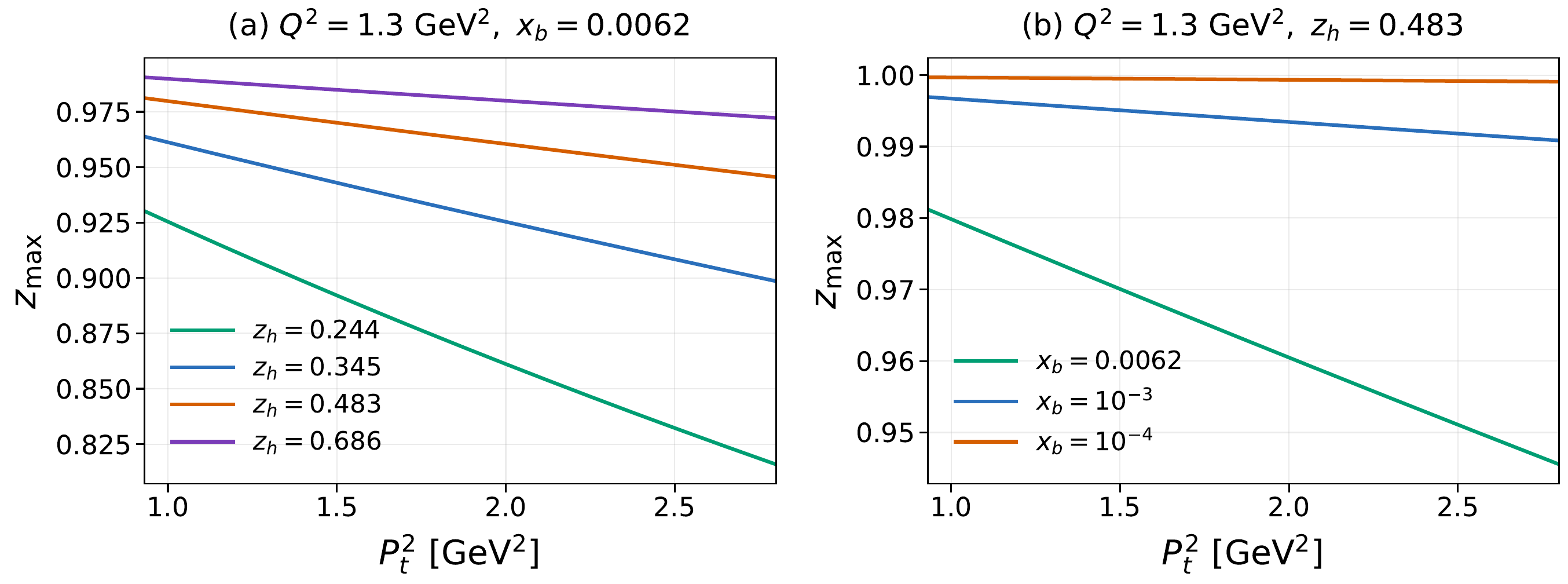}
    \caption{The variation of $z_{\rm max}$ (in Eq.\,\eqref{eq:y_max_and_xi}) with $P_t^2$ }
    \label{fig:y_max_variation}
\end{figure}

The NLO collinear curve provides the best description of the multiplicity. The $P_t$ fall-off of the CGC cross-section is much slower than in the COMPASS data. In order to understand this behavior, we start by comparing the limits of the quark longitudinal momentum fraction $z$ for the two cross-sections Eq.\,\eqref{eq:SIDIS_collinear} and Eq.\,\eqref{eq:SIDIS_small_x}. The upper limit $z_{\rm max}$ in the collinear cross-section is suppressed by momentum $P_t^2$, as can be seen in Fig.\,\ref{fig:y_max_variation}(a). However, the CGC cross-section has no such kinematic constraint on the upper limit $z_{\rm max}$. Therefore the suppression with $P_t^2$ is slower for the CGC in comparison to collinear factorization. We can also understand this slow suppression to be a result of working at an $x_B$ that is not low enough to ensure applicability of the CGC at large $P_t$. As we can see in Fig.\,\ref{fig:y_max_variation}(b), $z_{\rm max} \approx 1$ is a good approximation for a large range of $P_t$ as we go to small $x_B$. 

\begin{figure}[tbp]
  \begin{center}
  \includegraphics[width=0.85\linewidth]{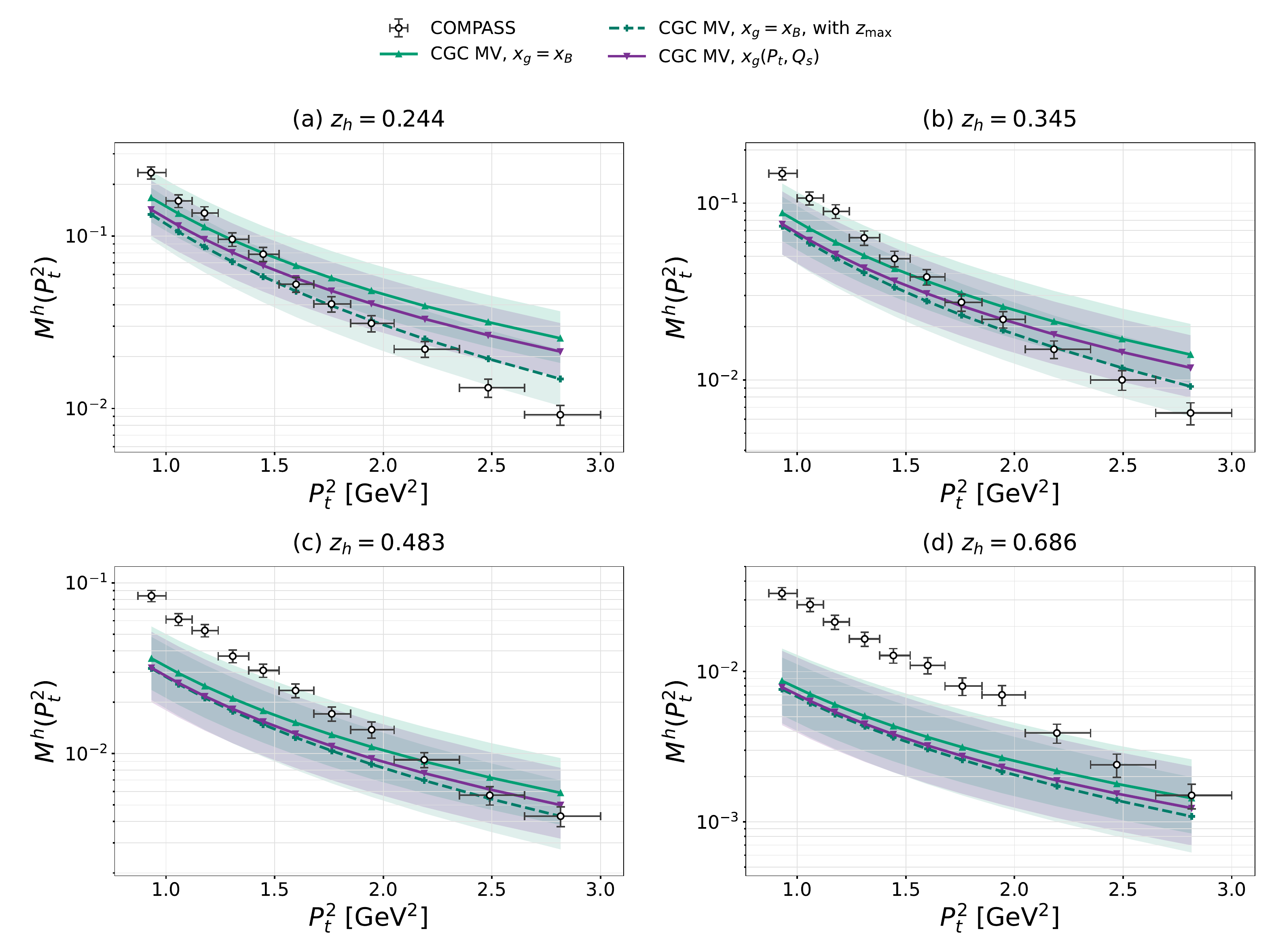}
  \end{center}
  \caption{Comparison to COMPASS data at $Q^2=1.3\,{\rm GeV}^2,\,x_B\,=\,0.0062$ and $y=0.69$. We show the original CGC cross-section Eq.\,\eqref{eq:SIDIS_small_x}, and a version with an upper bound $z_{\rm max}$ given in Eq.\,\eqref{eq:y_max_and_xi}, along with the CGC result using $x_g(P_t, Q_s)$ instead of $x_g=x_B$. The PDF, FF and uncertainty analysis is the same as in Fig.\,\ref{fig:multiplicity_COMPASS_without_ymax}.}
  \label{fig:multiplicity_COMPASS_with_ymax}
\end{figure}

To test the effect of this kinematic constraint on $z$ within the CGC calculation, we replace the upper limit of the $z$ integral in Eq.\,\eqref{eq:SIDIS_small_x} with $z_{\rm max}$ for COMPASS kinematics. Re-plotting the variation of multiplicity with $P_t^2$ under this approximation we get Fig.\,\ref{fig:multiplicity_COMPASS_with_ymax}, where we see that the agreement of the $P_t^2$ dependence of the CGC result with that of the experimental data has improved.

 We now analyze the effect of the scale choice $x_g = x_B$ on the $P_t$ dependence of our results. To this end, we consider an alternative $x_g$ motivated by longitudinal momentum conservation in the dijet production cross-section. In dijet production, imposing longitudinal momentum conservation fixes the momentum fraction $x_g$ to be
\begin{align}
\label{eq:x_g}
x_g=\,x_B \left(1 + \frac{k_1^2}{Q^2\,z} +  \frac{k_2^2}{Q^2\,(1-z)}\right)\,,
\end{align}
where $k_1$ and $k_2$ are the transverse momenta of the outgoing jets. To obtain SIDIS cross-section from the dijet production cross-section, we integrate over the antiquark momenta (or quark momenta), similar to Appendix \ref{app:dijet}. To get an analytic expression, however, we need to approximate $k_2^2$ with the dominant contribution. Following \cite{Iancu:2020jch}, we get
\begin{align}
\label{eq:P_t suppress}
x_g(P_t,Q_s)\,=\,x_B \left(1 + \frac{z P_t^2}{Q^2\,z_h^2} +  \frac{{\rm max}\left(\epsilon_f^2, Q_s^2(x_B)\right)}{Q^2\,(1-z)}\right)\,.
\end{align}
It should be noted that \cite{Iancu:2020jch} split the cross-section into elastic and inelastic pieces, where Eq.\,\eqref{eq:P_t suppress} is the choice for the in-elastic cross-section. Observing that the elastic piece contributes negligibly for the kinematics studied in this paper, we use Eq.\,\eqref{eq:P_t suppress} for the full SIDIS cross-section. For all $x_g(P_t,Q_s)$  exceeding the initial $x_0 = 0.01$ we use the dipole amplitude at $x_0$. \par 

As expected, in the strict high energy limit $x_B \rightarrow 0$, the different choices of $x_g$ are  equivalent. This is the regime where the longitudinal momentum transfer from the target is so small that longitudinal momentum conservation is satisfied trivially.  This is precisely where the eikonal CGC effective theory should be considered valid \cite{Brown:2026qlc}. Therefore, at non-zero $x_B$, Eq.\,\eqref{eq:P_t suppress} should be understood as modeling the effects of sub-eikonal corrections. 

Computing the SIDIS cross-section with Eq.\,\eqref{eq:P_t suppress} as the choice for $x_g(P_t, Q_s)$ leads to further $P_t$ suppression, as can be seen in Fig.\,\ref{fig:multiplicity_COMPASS_with_ymax}. 
The variations with the constraint $z_{\rm max} < 1$ and the $P_t$-dependent $x_g$ indicate that the LO eikonal-CGC is not sufficient to describe the large $P_t$ behavior for COMPASS kinematics. This motivates the inclusion of beyond-eikonal corrections \cite{Mukherjee:2026cte,Mukherjee:2026six,Kar:2026cnx,Altinoluk:2026SIDIS,Altinoluk:2023QuarkTMD} in the future. Since there is a notable jump in the collinear prediction from LO to NLO, it will also be interesting to compute the NLO CGC predictions in the future. Since COMPASS kinematics is close to projected EIC kinematics \cite{Accardi:2012qut}, we expect these conclusions to hold for the EIC as well.

\FloatBarrier

\section{HERA kinematics}\label{sec:HERA}

\begin{figure}[tbp]
  \begin{center}
  \includegraphics[width=\linewidth]{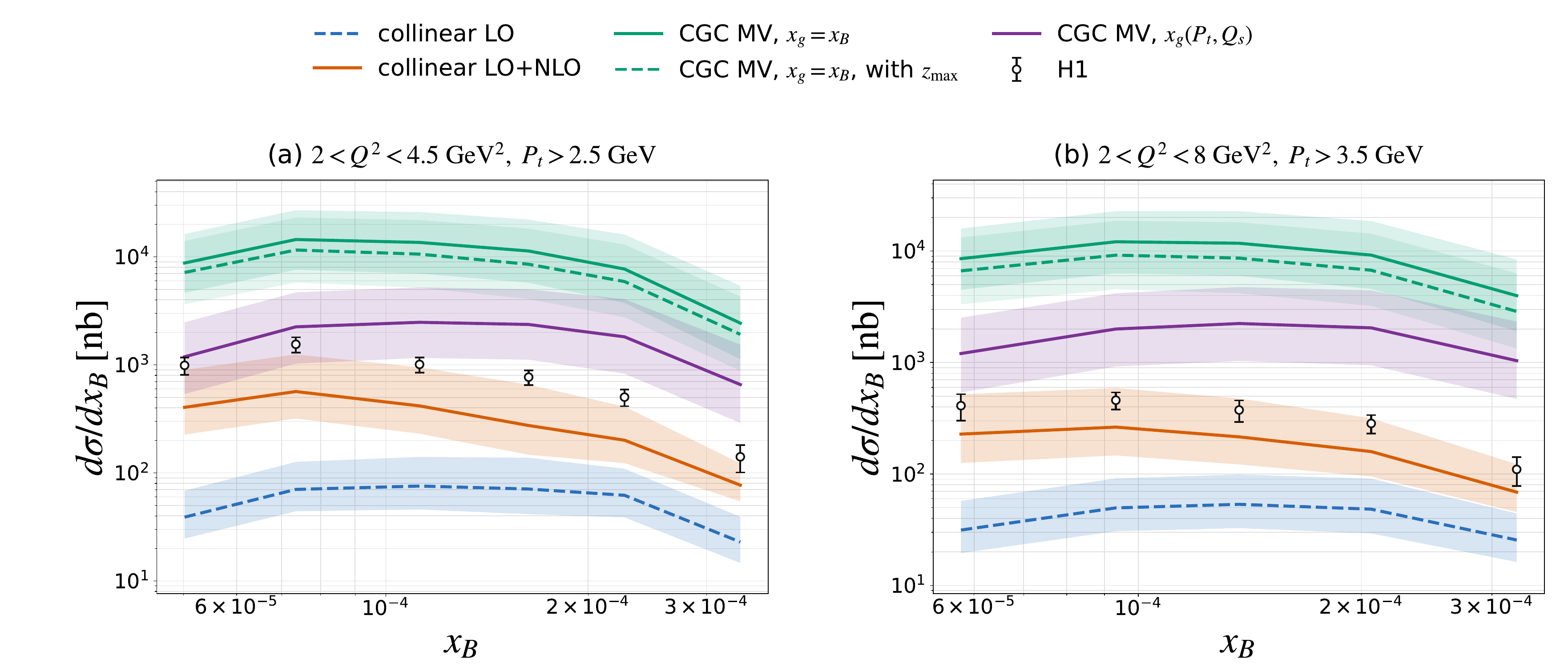}
  \end{center}
  \caption{The differential cross-section $d\sigma/dx_B$ vs $x_B$ obtained from the differential SIDIS cross-section by integrating over H1 cuts \cite{H1:2004xgw}. We use CT18NLO/CT18NNLO for PDF and NNFF10NLO/NNFF10NNLO for collinear LO/NLO plots. We use the MV model with NNFF10NLO for the CGC calculation. We observe a significant dependence on the kinematic constraints imposed on top of the CGC. Uncertainty bands for collinear plots are obtained by combining PDF,FF replica uncertainty and  scale variation $\mu = r Q$ with $r \in [1,4]$. The CGC uncertainty band is constructed from a combination of varying $Q_{s0}^2 \in [0.104,0.416]\,{\rm GeV^2}$ and FF scale variation by factors of 2 around its central value. The bold lines correspond to the choices $\mu = 2Q$ and $Q_{s0}^2 = 0.208$ GeV$^2$.}
  \label{fig:H1 vs x}
\end{figure}

In this section, we consider the H1 kinematics \cite{H1:2004xgw}. The H1 Collaboration measured the inclusive $\pi^0$-meson cross section $d\sigma_{\pi}/dx_B$ as a function of $x_B$ (Table 1 in \cite{H1:2004xgw}), integrating over $Q^2$, $P_t^2$, and $z_h$ in accordance with the H1 cuts detailed below. Since we are interested in the region $P_t \gtrsim Q$, we restrict ourselves to the first two columns of Table 1. The cross section is computed by imposing the HERA cuts on the ratio of the $\pi^0$ energy to the proton energy, $x_\pi=E_{\pi^0}/E_p$, the polar scattering angle relative to the proton direction, $\theta_\pi$, and inelasticity $y$:
\begin{align}
\label{eq:H1 cuts}
x_\pi \geq 0.01\,,\qquad 5^\circ<\theta_\pi<25^\circ\,,\qquad 0.1<y<0.6\,.
\end{align}
The choices of PDF \cite{Hou:2019efy} and FF \cite{Bertone:2017tyb} are the same as in the calculation for COMPASS kinematics. Since we do not compute a ratio to the DIS cross section here, we must choose a normalization (the transverse area of the proton) for the CGC cross section. We choose $\mathcal{A}_p=18.81$ mb, in accordance with the fit of the MV-model DIS cross section to HERA data \cite{Lappi:2013zma}. The cross section $d\sigma_{\pi}/dx_B$ as a function of $x_B$ is shown in Fig.\,\ref{fig:H1 vs x}.\par

\begin{figure}[tbp]
  \begin{center}
  \includegraphics[width=\linewidth]{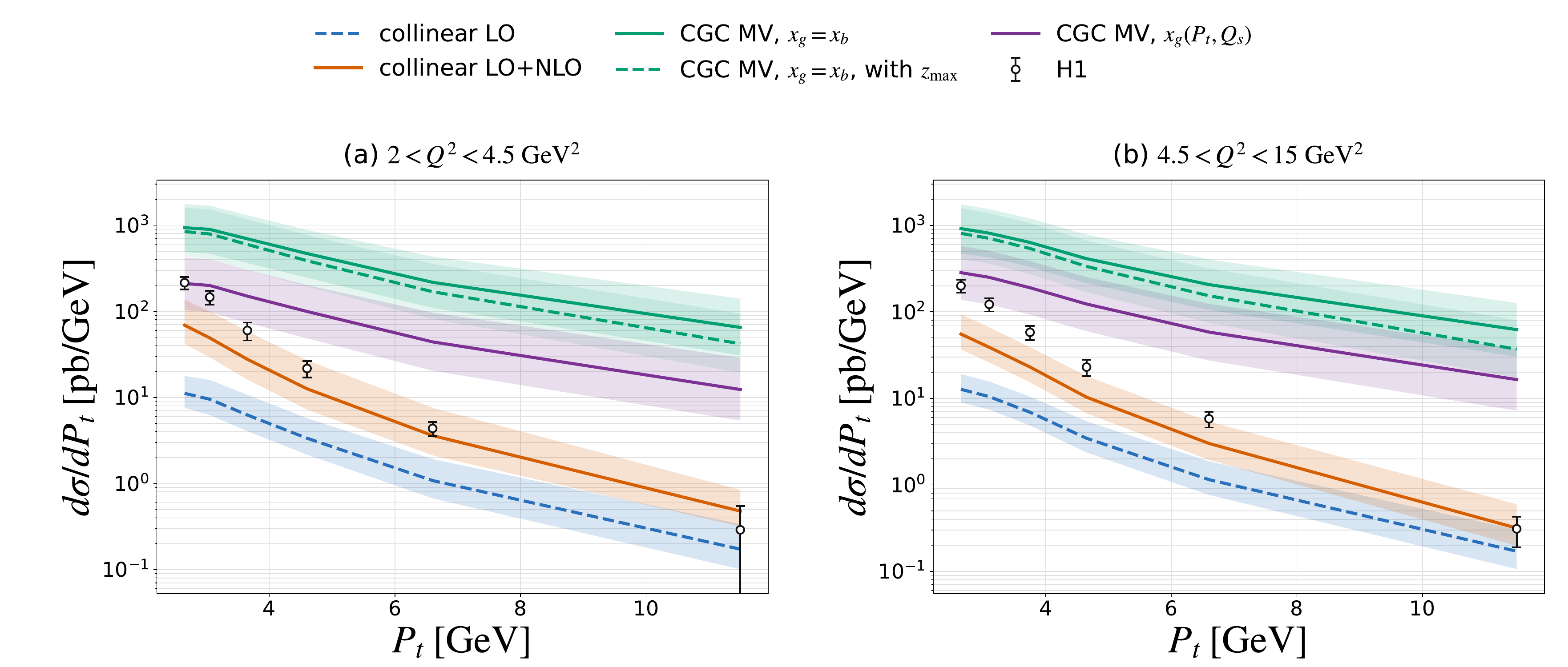}
  \end{center}
  \caption{The differential cross-section $d\sigma/dP_t$ vs $P_t$ obtained by integrating over H1 cuts \cite{H1:2004xgw}. We use the same method to generate the uncertainty bands as in Fig.~\ref{fig:H1 vs x}.}
  \label{fig:H1 vs pt}
\end{figure}

The H1 Collaboration \cite{H1:2004xgw} also measured the inclusive $\pi^0$-meson cross section $d\sigma_{\pi}/dP_t$ as a function of $P_t$ (Table 2 in \cite{H1:2004xgw}). Following the constraint $P_t \gtrsim Q$, we consider only the data in the first two columns of Table 2. We impose the H1 cuts in Eq.\,\eqref{eq:H1 cuts}, together with the HERA constraint $4\times10^{-5}\leq x_B\leq6\times10^{-3}$. The cross section $d\sigma_{\pi}/dP_t$ as a function of $P_t$ is shown in Fig.\,\ref{fig:H1 vs pt}. Collinear predictions for the H1 $d\sigma_{\pi}/dP_t$ cross section \cite{H1:2004xgw} were previously computed in Refs.\,\cite{Gonzalez-Hernandez:2018ipj,Daleo:2004pn}. We reproduce them in Fig.\,\ref{fig:H1 vs pt}, together with the CGC results. The small differences to the results of \cite{Gonzalez-Hernandez:2018ipj,Daleo:2004pn} can be attributed to a different choice of PDF and FF.

Although the collinear NLO calculation provides the best description, we note that the normalization of the CGC curve with $x_g=x_B$ (green) is much higher than the H1 results in Figs.\,\ref{fig:H1 vs x} and \ref{fig:H1 vs pt}. For comparison, using $x_g(P_t,Q_s)$ as defined in Eq.\,\eqref{eq:P_t suppress} reduces the overall normalization of the CGC curve, bringing it closer to the H1 data and the collinear NLO predictions (see the purple curves in Figs.\,\ref{fig:H1 vs x} and \ref{fig:H1 vs pt}). This again demonstrates the sensitivity of the LO eikonal SIDIS cross section to the choice of $x_g$ and, consequently, its failure to capture the $P_t$ dependence in HERA kinematics. The constraints associated with $z_{\rm max}$ and $x_g(P_t,Q_s)$ signal the need to incorporate beyond-eikonal effects \cite{Mukherjee:2026cte,Mukherjee:2026six,Kar:2026cnx,Altinoluk:2026SIDIS,Altinoluk:2023QuarkTMD}, especially at large $P_t$. This is evident in Fig.\,\ref{fig:H1 vs pt}, where the mismatch increases with $P_t$. 

Separately, we observe that the scale variation of the NLO collinear curve is larger than that of the LO collinear curve. Since the NLO correction is large, this scale variation suggests that perturbative convergence may not yet have been reached for H1 kinematics. NLO effects can therefore also be expected to affect the CGC predictions.

\begin{figure}[tbp]
  \begin{center}
\includegraphics[width=\linewidth]{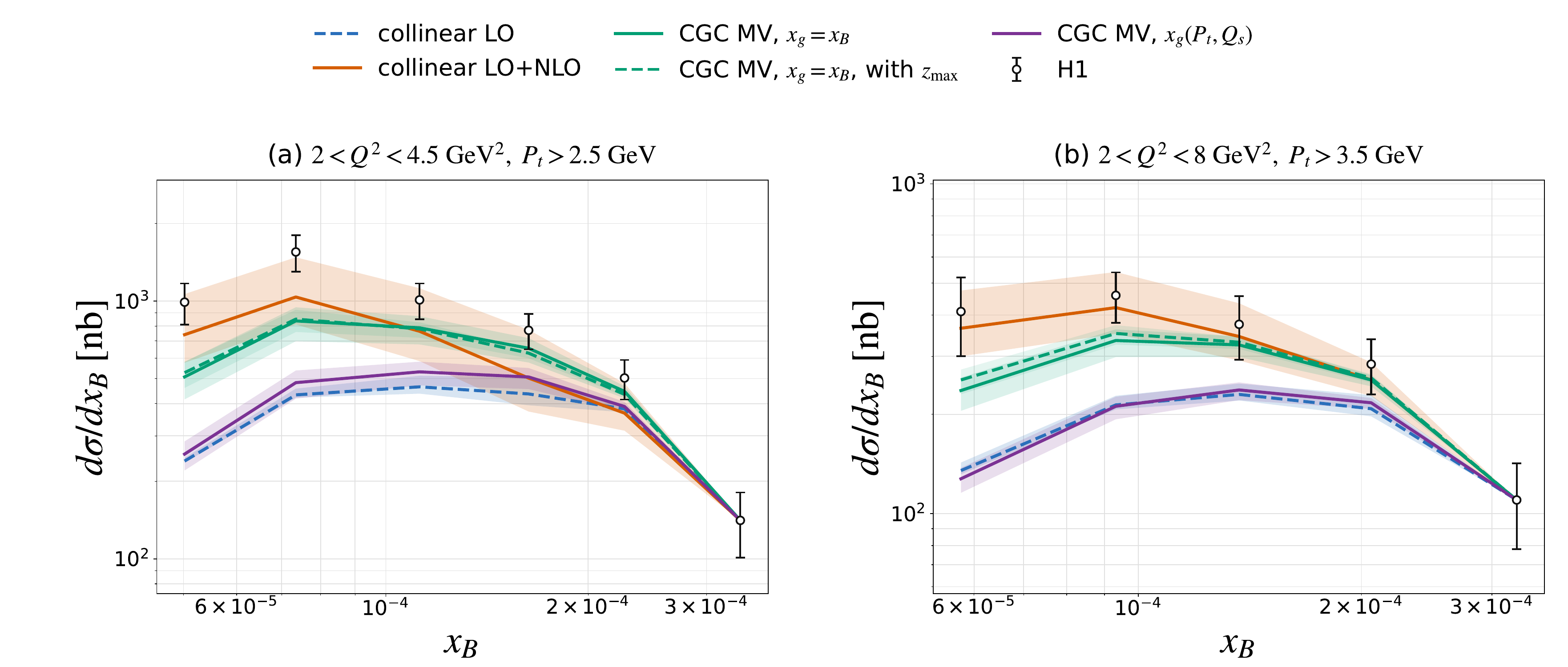}
  \end{center}
  \caption{The normalized differential cross-section $d\sigma/dx_B$ vs $x_B$ obtained by integrating over H1 cuts \cite{H1:2004xgw} and scaling all curves to agree at the largest value of $x_B$ available in the experimental data. We use the same method to generate the uncertainty bands as Fig.~\ref{fig:H1 vs x}.}
  \label{fig:H1 vs x normalized}
\end{figure}

To test the $x_B$ dependence independently of the normalization, we normalize all four curves to their values at the largest available $x_B$ point in Fig.\,\ref{fig:H1 vs x normalized}. We find that the collinear NLO calculation provides the best description of the $x_B$ dependence. The LO CGC curve with $x_g=x_B$ captures the $x_B$ dependence better than the collinear LO curve. We also note that the purple curve obtained using $x_g(P_t,Q_s)$ appears to have approximately the same $x_B$ dependence as the collinear LO curve.

\FloatBarrier

\section{SIDIS with nuclear targets}\label{sec:nuclear}
In this section we analyze the nuclear SIDIS cross-section $e + N\rightarrow h + X$, where we take the outgoing hadron to be positively charged, as done by COMPASS. Since no data is available for large nuclei, we make predictions within the collinear factorization and the CGC frameworks, using the cross-sections Eq.\,\eqref{eq:SIDIS_collinear} and Eq.\,\eqref{eq:SIDIS_small_x}, respectively. Concretely, we predict the ratio between the cross section with a nuclear target and the scaled cross section for a proton target
\begin{align}
\label{eq:def_ratio_nuclear}
    \frac{d \sigma_N}{A\,d\sigma_p}\,=\,\frac{1}{A}\frac{d \sigma^{e N\:\rightarrow\:ehX}}{dQ^2 d x_B d z_h dP_t^2}\bigg/\frac{d \sigma^{e p\:\rightarrow\:ehX}}{dQ^2 d x_B d z_h dP_t^2}\,,
\end{align}
where $A$ is the mass-number of the nucleus under consideration. Note that we have suppressed the explicit differentials on the left for brevity. 

In the case of collinear factorization, we evaluate the nuclear SIDIS cross-section by employing a nuclear PDF \cite{Kovarik:2015cma,Duwentaster:2022kpv,AbdulKhalek:2019mzd,
AbdulKhalek:2020yuc,Eskola:2021nhw,Walt:2019slu,Helenius:2021tof,
Hou:2019efy}. Since nuclear PDFs are normalized per-nucleon, we do not require the explicit division by $A$ in Eq.\,\eqref{eq:def_ratio_nuclear} in practice. Nuclear PDFs involve baseline proton PDFs, and we make sure to use the appropriate proton PDF in the SIDIS cross section for proton targets in the denominator of Eq.\,\eqref{eq:def_ratio_nuclear}. 

In the CGC case, we define the nuclear SIDIS cross-section with parameters $Q_{sN}^2 = A^{1/3} Q_{s0}^2$ and $\mathcal{A}_N\,=\,A^{2/3}\mathcal{A}_p$. We motivate these choices by the expectation that, in the large $P_t$ limit, the per-nucleon properties of the nucleus are equivalent to the properties of the proton, thus making the ratio in Eq.\,\eqref{eq:def_ratio_nuclear} approach 1. We note that in a more sophisticated geometric model, the $A$ dependence of $Q_{sN}^2$ and the area $\mathcal{A}_N$ may vary \cite{Deganutti:2023qct}.

Some of the nuclear PDFs are available at NLO  \cite{Kovarik:2015cma,Eskola:2021nhw,Duwentaster:2022kpv,AbdulKhalek:2020yuc} while some are given at NNLO \cite{AbdulKhalek:2019mzd,Walt:2019slu,Helenius:2021tof}. We use NNFF10NLO (see \cite{Bertone:2017tyb}) for nuclear PDFs available at NLO and we use NNFF10NNLO for nuclear PDFs at NNLO. We use the collinear hard factor at NLO order for all nuclear predictions. For the CGC, we choose the FF to be NNFFNLO. We perform this analysis for anticipated EIC kinematics: $x_B = 5 \times 10^{-4}$, $Q^2 = P_t^2 = 4\,{\rm GeV}^2$, and inelasticity $y = 0.69$. 

We show the ratio Eq.\,\eqref{eq:def_ratio_nuclear} as a function of mass number $A$ for EIC kinematics with $z_h = 0.5$ in Fig.\,\ref{fig:a_dependence_SIDIS} and in Fig.\,\ref{fig:a_dependence_SIDIS_abnormal} (For other $z_h$ values see Appendix \ref{app:Ratio_other_zh}). We include the nuclear PDF sets nCTEQ15 \cite{Kovarik:2015cma}, EPPS21 \cite{Eskola:2021nhw}, Tuju19 \cite{Walt:2019slu} and Tuju21 \cite{Helenius:2021tof}. There are additional nPDF sets nCTEQ15HQ\cite{Duwentaster:2022kpv}, NNPDF1.0 \cite{AbdulKhalek:2019mzd} and nNNPDF2.0 \cite{AbdulKhalek:2020yuc}, which we separate from Fig.\,\ref{fig:a_dependence_SIDIS} since they have an unexpected rise in the ratio for large $A$ (see Fig.\,\ref{fig:a_dependence_SIDIS_abnormal}). Note that the EPPS21 result shows unusually little suppression for $^6$Li nuclei, which might have a distinct $\alpha+d$ structure. This behavior is not reproduced by the other nPDFs because the way the $A$ dependence is determined smooths over outliers like this. 

In Figures \,\ref{fig:a_dependence_SIDIS} and \ref{fig:a_dependence_SIDIS_abnormal} we see that for the CGC result using $x_g = x_B$ the presented ratio is suppressed compared to the typical results from collinear factorization. Compared to the CGC result using $x_g = x_B$, the result from the CGC with $x_g(P_t,Q_s)$ shows less suppression. The observed variation with the choice of $x_g$ tells us that $x_B$ for EIC kinematics is not small enough for a robust eikonal CGC prediction. 

The large errors associated with the nuclear PDFs also do not allow for a clean conclusion. Since the nuclear PDFs are expected to be further constrained by the time of arrival of the EIC SIDIS data, it will allow us to make a stronger statement about the comparison of collinear factorization and the CGC in the future. In anticipation of early science runs at the EIC, we also show the silver nucleus to proton ratio vs $P_t^2$ for EIC kinematics in Fig.\,\ref{fig:silver_EIC}. We again observe a large variation of the CGC prediction with the choice of $x_g$, with more suppression for $x_g=x_B$. 
For the $x_g(P_t,Q_s)$ prescription, the nuclear ratio can exceed unity at small values of $z_h$, as seen in Fig.\,\ref{fig:silver_EIC}. This behavior can be traced to the fact that $x_g(P_t,Q_s)$ is shifted to values larger than $x_B$, with the shift becoming more pronounced at smaller $z_h$. Consequently, the dipole amplitude is probed at larger $x_g$, where saturation effects are weaker due to lack of rcBK evolution, thereby reducing the nuclear suppression and, for smaller values of $z_h$, leading to a ratio above unity.
 \par

\begin{figure}[tbp]
  \begin{center}
  \includegraphics[width=0.82\linewidth]{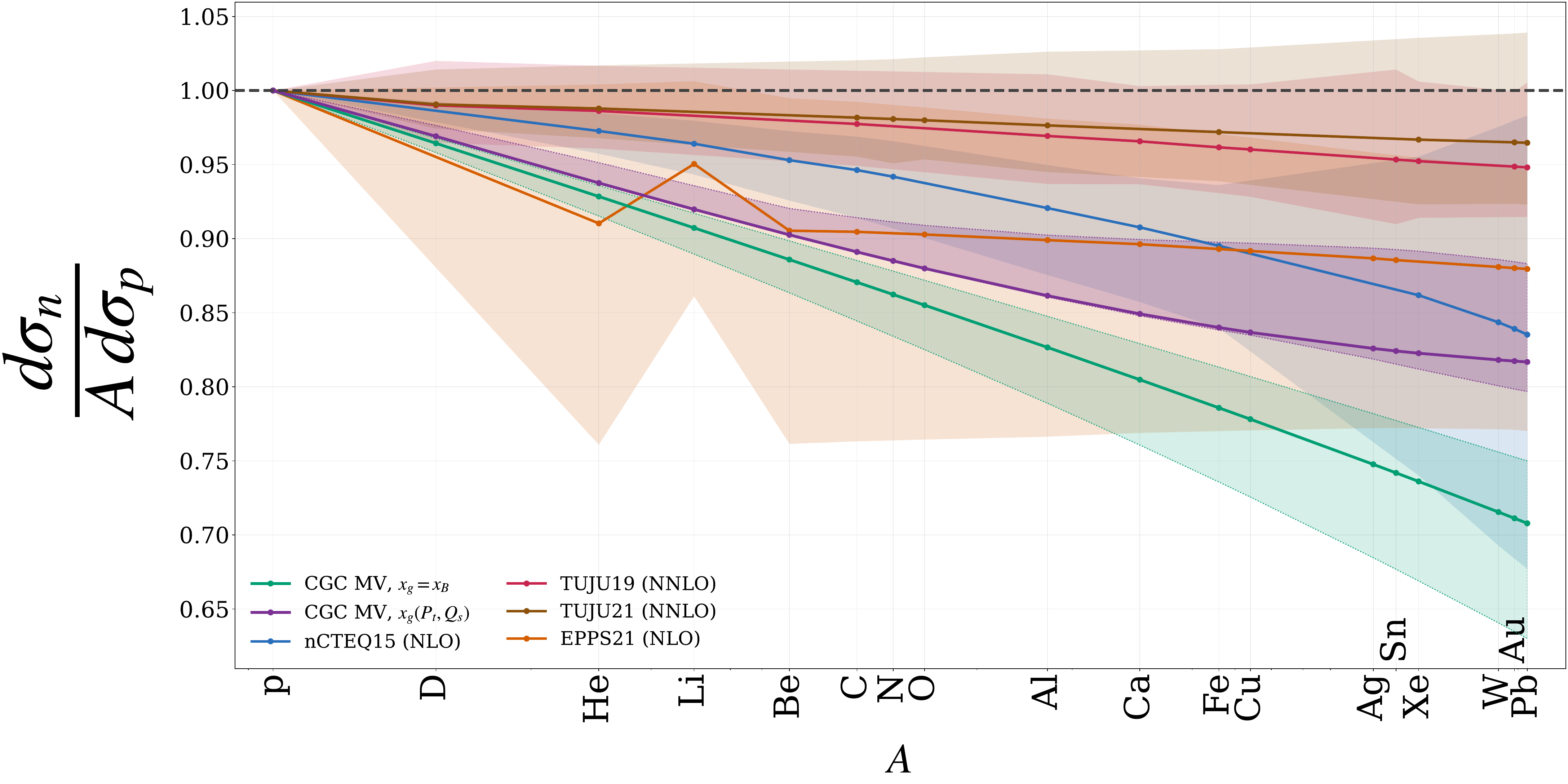}
  \end{center}
  \caption{Variation of ratio Eq.\,\eqref{eq:def_ratio_nuclear} with $A$ at EIC kinematics $x_B = 5\times10^{-4},\,z_h=0.5,\,Q^2 = P_t^2=  4\,{\rm GeV}^2$ and $y= 0.69$ for the nPDF sets nCTEQ15 \cite{Kovarik:2015cma}, EPPS21 \cite{Eskola:2021nhw}, TUJU19 \cite{Walt:2019slu} and TUJU21 \cite{Helenius:2021tof}. Uncertainty bands for collinear plots are obtained by combining nPDF,FF replica uncertainty and  scale variation $\mu = r Q$ with $r \in [1,4]$. The CGC uncertainty band is constructed from a combination of $Q_{s0}^2 \in [0.104,0.416]\,{\rm GeV^2}$ and FF scale variation. The bold lines correspond to the choices $\mu = 2Q$ and $Q_{s0}^2 = 0.208$ GeV$^2$. Here, NLO and NNLO denote the highest available perturbative order in \(\alpha_s\) for the corresponding nuclear PDF sets.}
  \label{fig:a_dependence_SIDIS}
\end{figure}

\begin{figure}[tbp]
  \begin{center}
  \includegraphics[width=0.82\linewidth]{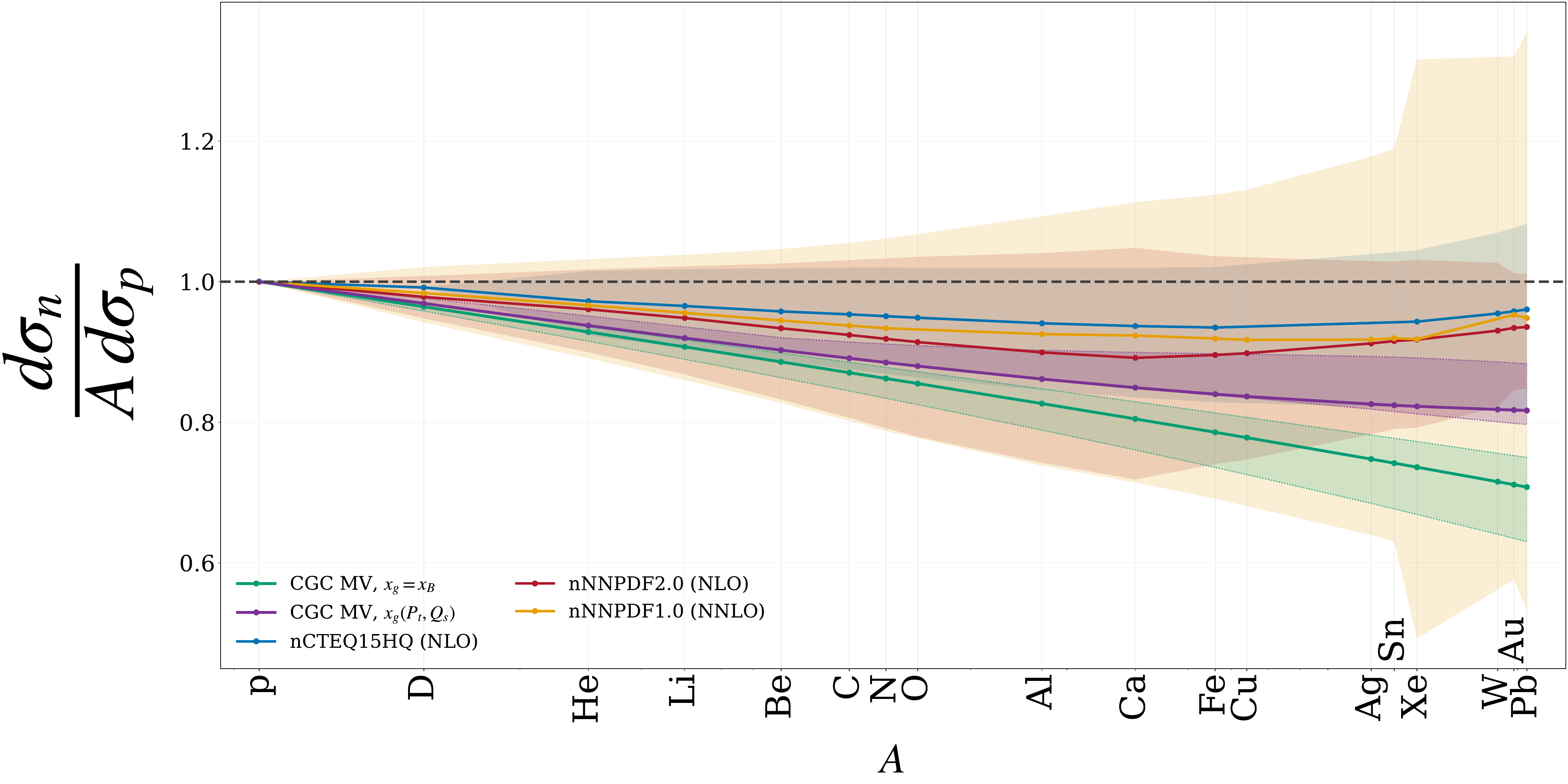}
  \end{center}
  \caption{Variation of ratio Eq.\,\eqref{eq:def_ratio_nuclear} with $A$ at EIC kinematics $x_B =5 \times 10^{-4},\,z_h=0.5, Q^2 =P_t^2= 4\,{\rm GeV}^2$ and $y= 0.69$ for nNNPDF1.0 \cite{AbdulKhalek:2019mzd}, nNNPDF2.0 \cite{AbdulKhalek:2020yuc} and nCTEQ15HQ \cite{Duwentaster:2022kpv}. We note that nNNPDF1.0, nNNPDF2.0 and nCTEQ15HQ and sets show an increase of the ratio with mass number A. The uncertainty bands are computed similar to Fig\,\ref{fig:a_dependence_SIDIS}.}
  \label{fig:a_dependence_SIDIS_abnormal}
\end{figure}

\begin{figure}[tbp]
  \begin{center}
  \includegraphics[width=0.82\linewidth]{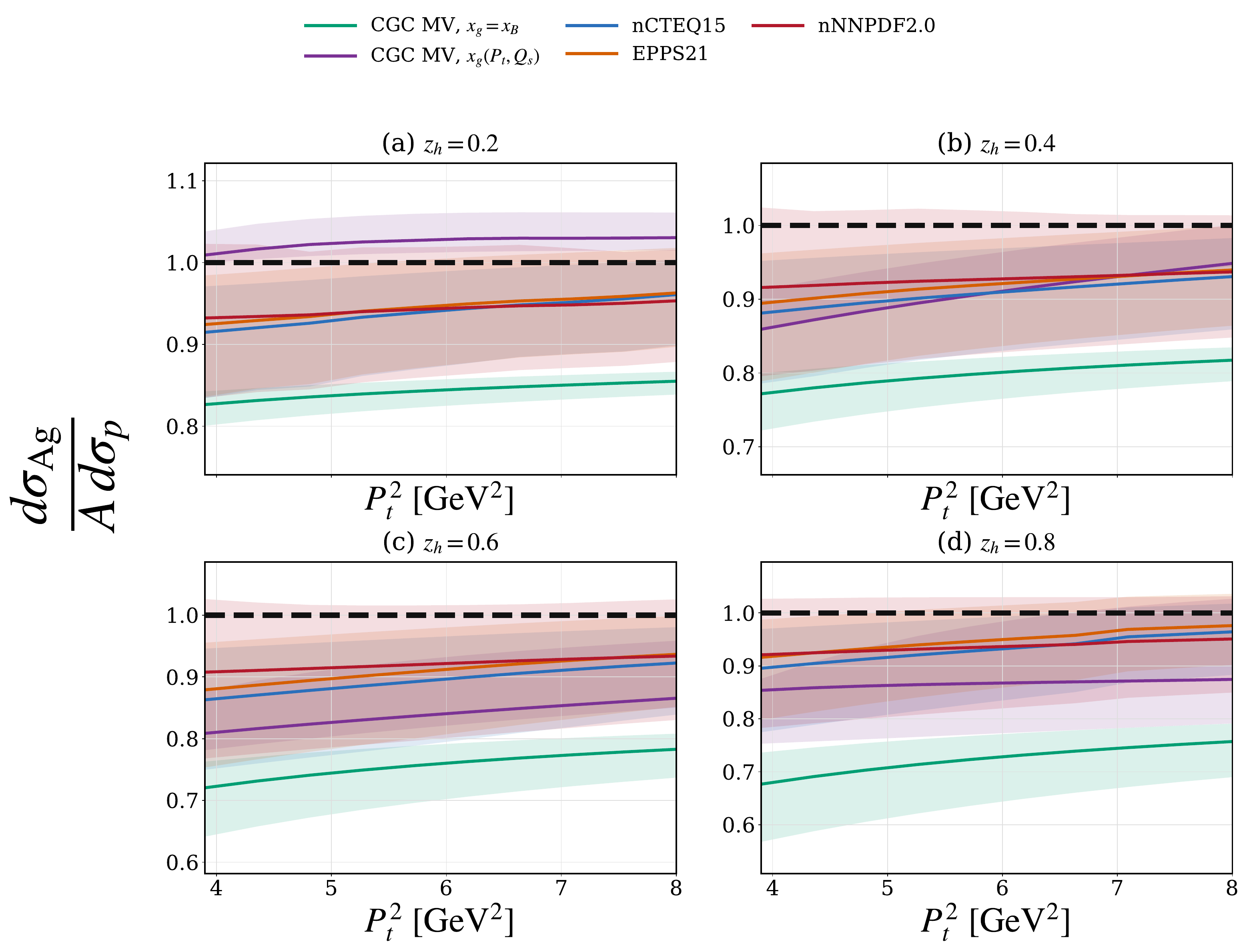}
  \end{center}
  \caption{Transverse momentum dependent silver nucleus (A = 108) to proton SIDIS cross section ratio Eq.\,\eqref{eq:def_ratio_nuclear} for EIC kinematics $x_B = 5\times 10^{-4},\,Q^2 = 4\,{\rm GeV}^2$ and $y= 0.69$ for the CGC case with two different choices for $x_g$ and for the collinear case using nNNPDF2.0 \cite{AbdulKhalek:2020yuc}, EPPS21 \cite{Eskola:2021nhw} and nCTEQ15 \cite{Kovarik:2015cma}. The uncertainty bands are computed as in Fig.\,\ref{fig:a_dependence_SIDIS}.}
  \label{fig:silver_EIC}
\end{figure}
\FloatBarrier
\section{Conclusions}\label{sec:conclusion}
We performed a systematic study of large-$P_t$ SIDIS cross-sections using both the collinear factorization framework and CGC effective theory. We use up to NLO SIDIS cross-sections in the collinear framework while restricting ourselves to the LO in the CGC framework. We leave the NLO SIDIS calculation in the CGC framework for a future study. The resulting cross-section predictions are then compared with the measurements at COMPASS and HERA.

Starting with COMPASS \cite{COMPASS:2017mvk}, we find that the ratio $M^h(P_t)$ between the SIDIS cross-section and the DIS cross-section is best described by NLO collinear prediction. The $P_t$ fall-off of the CGC cross-section does not fully reproduce the  COMPASS data at $x_B\,=\,0.0062$. This remains true for HERA H1 \cite{H1:2004xgw}  (with $4 \times 10^{-5} \leq x_B \leq 6 \times 10^{-3}$), which involves measurements at a much larger $P_t$. This indicates that LO-eikonal CGC is unable to describe the $P_t$ fall-off for HERA and COMPASS kinematics. This is because LO-eikonal CGC works well when the longitudinal momentum transfer from the target is small. However, for large values of $P_t$, this momentum transfer can be significant. Thus, one requires smaller values of $x_B$ to ensure the validity of CGC for a larger range of $P_t$. As a further test, we implement effects of longitudinal momentum conservation in the CGC formalism and get improved behavior with $P_t$ by making the momentum fraction $x_g$ a function of $P_t$.  This variation with the choice of $x_g$ signals a need for inclusion of beyond-eikonal corrections to the CGC SIDIS predictions for EIC kinematics. Furthermore, we observe that the collinear predictions have a large NLO correction. This also motivates the inclusion of NLO CGC corrections in the future.

We finally studied the ratio between the nuclear SIDIS cross-section and the proton SIDIS cross-section in the two factorization schemes. We implemented nuclear effects by introducing nuclear PDFs for collinear factorization and by modifying the saturation scale $Q_{sN}^2\,=\,A^{1/3} Q_{s0}^2$ for the CGC cross-section, with $A$ the mass number. We again observed that the CGC cross-section shows a large variation with the choice of $x_g$ for EIC kinematics. It will be important to understand if this variation will be reduced by the inclusion of beyond-eikonal and NLO corrections to the CGC in the future. 

\section{Acknowledgments}
We thank Heikki M\"antysaari, Swaleha Mulani, Farid Salazar, Vladi Skokov and Fei Yao for illuminating discussions. This work is supported by the U.S. Department of Energy, Office of Science, Office of Nuclear Physics, under DOE Contract No.~DE-SC0012704 and within the framework of the Saturated Glue (SURGE) Topical Theory Collaboration.

\appendix
\section{Conventions}
\label{app:Conv}
The relations between various kinematic variables used in the main text are given by
\begin{equation}
\begin{aligned}
q\:=\:l-l',\:Q^2=\:-q^2,\:x_B=& {Q^2 \over 2 P_N.q}, z_h\:=\:{P_N.P_h \over P_N.q},W^2 = -Q^2\left(1 - \frac{1}{x_B}\right)\,,\,y={P_N.q \over P_N.l},\na &\:s\,=\,(P_N+l)^2\,,Q^2\:=\:x_Bys\,. \,
\end{aligned}
\end{equation}
Here $s$ is the center of mass energy squared of the process and the momenta $P_N,q,l$ and $l'$ are defined in figure \ref{fig:SIDIS}. We note that $z_h$ and $P_t$ correspond to the longitudinal momentum fraction and transverse momentum of the measured hadron. We neglect the target mass in this calculation. 

\section{SIDIS cross-section from dijet production}
\label{app:dijet}
In this appendix, we show how to derive Eq.\,\eqref{eq:SIDIS_partonic_CGC} in the CGC from the leading-order dijet production cross-section. We start from the LO dijet expression for longitudinal and transverse polarizations \cite{Caucal:2023fsf,Iancu:2020jch,Dominguez:2011wm,Dominguez:2011br,Caucal:2021ent}
\begin{align}
&\frac{d \sigma^{\gamma^*_{\rm L} N \rightarrow q\bar q X}}{dk_1^-\,d^2 \mathbf{k}_{1}\,dk_2^-\,d^2\mathbf{k}_{2}}\,=\,8\,(2\pi)\,\frac{N_c \alpha_{\rm em} e_q^2}{q^-}\,\bar \delta(q^- - k_1^- - k_2^-)\,z(1-z)\,\epsilon_f^2\,\int \dhd^2 \mathbf{x}_1\,\dhd^2 \mathbf{y}_1\,\dhd^2 \mathbf{x}_2\,\dhd^2 \mathbf{y}_2\,e^{-i \mathbf{k}_1.(\mathbf{x}_1 - \mathbf{y}_1)}\,\na &e^{-i \mathbf{k}_2.(\mathbf{x}_2 - \mathbf{y}_2)}\,K_0(\epsilon_f r_x)\,K_0(\epsilon_fr_y)\,\Big(1+ Q(\mathbf{x}_1,\mathbf{x}_2;\mathbf{y}_2,\mathbf{y}_1) - S(\mathbf{x}_1, \mathbf{x}_2) - S(\mathbf{y}_1, \mathbf{y}_2)\Big)\,, \na &
\frac{d \sigma^{\gamma^*_{\rm T} N \rightarrow q\bar q X}}{dk_1^-\,d^2 \mathbf{k}_{1}\,dk_2^-\,d^2\mathbf{k}_{2}}\,=\,2\,(2\pi)\,\frac{N_c \alpha_{\rm em} e_q^2}{q^-}\,\bar \delta(q^- - k_1^- - k_2^-)\,\left(z^2 + (1-z)^2\right)\,\epsilon_f^2\,\int \dhd^2 \mathbf{x}_1\,\dhd^2 \mathbf{y}_1\,\dhd^2 \mathbf{x}_2\,\dhd^2 \mathbf{y}_2\,\na &e^{-i \mathbf{k}_1.(\mathbf{x}_1 - \mathbf{y}_1)}\,e^{-i \mathbf{k}_2.(\mathbf{x}_2 - \mathbf{y}_2)}\,\frac{\mathbf{r_x}.\mathbf{r_y}}{r_x\,r_y}\,K_1(\epsilon_f r_x)\,K_1(\epsilon_fr_y)\,\Big(1+ Q(\mathbf{x}_1,\mathbf{x}_2;\mathbf{y}_2,\mathbf{y}_1) - S(\mathbf{x}_1, \mathbf{x}_2) - S(\mathbf{y}_1, \mathbf{y}_2)\Big)\,,
\end{align}
where $\bar \delta(x)\,=\,2\pi\,\delta(x)$ and $\dhd^2 \mathbf{x}\,=\,d^2\mathbf{x}/(2\pi)^2$. Here $\mathbf{k}_1$ and $\mathbf{k}_2$ are the momenta of the outgoing quark and anti-quark jets respectively. We have defined $\mathbf{r_x}\,=\,\mathbf{x}_1 - \mathbf{x}_2$ and $\mathbf{r_y}\,=\,\mathbf{y}_1 - \mathbf{y}_2$ along with $k_1^-\,=\, z\,q^-$ and $k_2^-\,=\, (1-z)\, q^-$. The function $Q(\mathbf{x}_1,\mathbf{x}_2;\mathbf{y}_1,\mathbf{y}_2)$ is the quadrupole amplitude
\begin{align}
\label{eq:Quadrupole}
Q(\mathbf{x}_1,\mathbf{x}_2;\mathbf{y}_2,\mathbf{y}_1)\,=\,\frac{1}{N_c}\,{\rm Tr}\left(V_{\mathbf{x}_1}\,V^{\dagger}_{\mathbf{x}_2}\,V_{\mathbf{y}_2}\,V^{\dagger}_{\mathbf{y}_1}\right)\,.
\end{align}
The dipole amplitude $S(\mathbf{x}_1, \mathbf{x}_2)$ is defined in Eq.\,\eqref{eq:dipole_amplitude}. In order to compute the SIDIS amplitude, we now integrate over the anti-quark momenta $k_2^-, \mathbf{k}_{2}$ (One can also integrate over the quarks). The integral over $k_2^-$ is trivial due to the delta function, while the integral over $\mathbf{k}_2$ produces a delta function $\delta(\mathbf{x}_2 - \mathbf{y}_2)$. Evaluating this delta function to remove the $\mathbf{y}_2$ integral, we see that the Eq.\,\eqref{eq:Quadrupole} reduces to the dipole Eq.\,\eqref{eq:dipole_amplitude}. Thus this gives us
\begin{align}
&\frac{d \sigma^{\gamma^*_{\rm L} N \rightarrow q\bar q X}}{dk_1^-\,d^2 \mathbf{k}_{1}}\,=\,8\,(2\pi)^2\,\frac{N_c \alpha_{\rm em} e_q^2}{q^-}\,z(1-z)\,\epsilon_f^2\,\int \dhd^2 \mathbf{x}_1\,\dhd^2 \mathbf{y}_1\,\dhd^2 \mathbf{x}_2\,e^{-i \mathbf{k}_1.(\mathbf{x}_1 - \mathbf{y}_1)}\,\na &\,K_0(\epsilon_f r_x)\,K_0(\epsilon_fr_y)\,\Big(1+ S(\mathbf{x}_1, \mathbf{y}_1) - S(\mathbf{x}_1, \mathbf{x}_2) - S(\mathbf{y}_1, \mathbf{x}_2)\Big)\,, \na &
\frac{d \sigma^{\gamma^*_{\rm T} N \rightarrow q\bar q X}}{dk_1^-\,d^2 \mathbf{k}_{1}}\,=\,2\,(2\pi)^2\,\frac{N_c \alpha_{\rm em} e_q^2}{q^-}\,\,\left(z^2 + (1-z)^2\right)\,\epsilon_f^2\,\int \dhd^2 \mathbf{x}_1\,\dhd^2 \mathbf{y}_1\,\dhd^2 \mathbf{x}_2\,\na &e^{-i \mathbf{k}_1.(\mathbf{x}_1 - \mathbf{y}_1)}\,\frac{\mathbf{r_x}.\mathbf{r_y}}{r_x\,r_y}\,K_1(\epsilon_f r_x)\,K_1(\epsilon_fr_y)\,\Big(1+ S(\mathbf{x}_1, \mathbf{y}_1) - S(\mathbf{x}_1, \mathbf{x}_2) - S(\mathbf{y}_1, \mathbf{x}_2)\Big)\,,
\end{align}
with $\mathbf{r_y}\,=\,\mathbf{y}_1 - \mathbf{x}_2$. We now show how to massage the above equation into the form Eq.\,\eqref{eq:SIDIS_partonic_CGC}. We shall explicitly show this for the longitudinal polarization. Let us start by re-writing the above expression in terms of $N(\mathbf{x}_1, \mathbf{x}_2)= 1 - S(\mathbf{x}_1, \mathbf{x}_2)$,
\begin{align}
&\frac{d \sigma^{\gamma^*_{\rm L} N \rightarrow q\bar q X}}{dk_1^-\,d^2 \mathbf{k}_{1}}\,=\,8\,\,\frac{N_c \alpha_{\rm em} e_q^2}{q^-}\,z(1-z)\,\epsilon_f^2\,\int d^2 \mathbf{R}\,\dhd^2 \mathbf{r}_1\,\dhd^2 \mathbf{r}_2\,e^{-i \mathbf{k}_1.(\mathbf{r}_1-\mathbf{r}_2)}\,\,K_0(\epsilon_f r_1)\,K_0(\epsilon_fr_2)\na &\,\Big(N(\mathbf{r}_1,\mathbf{R}) + N(-\mathbf{r}_2,\mathbf{R}) - N(\mathbf{r}_1 - \mathbf{r}_2,\mathbf{R})\Big)\,,
\end{align}
We can now change variables to write the above expression as
\begin{align}
&\frac{d \sigma^{\gamma^*_{\rm L} N \rightarrow q\bar q X}}{dk_1^-\,d^2 \mathbf{k}_{1}}\,=\,8\,(2\pi)^2\,\frac{N_c \alpha_{\rm em} e_q^2}{q^-}\,z(1-z)\,\epsilon_f^2\,\int d^2 \mathbf{R}\,\dhd^2 \mathbf{r}_1\,\dhd^2 \mathbf{r}_2\,\dhd^2 q_1\,\dhd^2 q_2\,\,\frac{1}{q_1^2 + \epsilon_f^2}\,\frac{1}{q_{2}^2 + \epsilon_f^2}\na &\,\Big(e^{i (\mathbf{q_1} - \mathbf{k}_1).\mathbf{r_1}}\,\,e^{i (\mathbf{q_2}+ \mathbf{k}_1).\mathbf{r_2}} +\,e^{-i (\mathbf{q_2} - \mathbf{k}_1).\mathbf{r_1}}\,e^{-i (\mathbf{q_1} +\mathbf{k}_1).\mathbf{r_2}}\, - \,e^{i (\mathbf{q_1}-\mathbf{k}_1).\mathbf{r_1}}\,\,e^{i (\mathbf{q_2}+\mathbf{q_1}).\mathbf{r_2}}\Big)N(\mathbf{r}_1,\mathbf{R})\,.
\end{align}
We now integrate over $\mathbf{r_2}$. This generates a Dirac delta function and allows us to integrate over momenta $\mathbf{q}_1$. Rewriting the other integral over $\mathbf{q}_2$ in terms of Bessel-K function, we get
\begin{align}
\frac{d \sigma^{\gamma^*_{\rm L} N \rightarrow q\bar q X}}{dk_1^-\,d^2 \mathbf{k}_{1}}\,&=\,8\,\frac{N_c \alpha_{\rm em} e_q^2}{(2\pi)\,q^-}\,z(1-z)\,\epsilon_f^2\,\int d^2 \mathbf{R}\,\dhd^2 \mathbf{r}_1\,\Big(K_0(\epsilon_f r_1)\,\frac{1}{k_1^2 + \epsilon_f^2}e^{-i  \mathbf{k}_1.\mathbf{r_1}}\, +\,K_0(\epsilon_f r_1)\,\frac{1}{k_1^2 + \epsilon_f^2}\,\,e^{i \mathbf{k}_1.\mathbf{r_1}} \na & - \,\frac{r_1}{2\epsilon_f}K_1(\epsilon_f r_1)\,e^{-i \mathbf{k}_1.\mathbf{r_1}}\,\Big)N(\mathbf{r}_1,\mathbf{R})\,.
\end{align}
We now re-cast the above differential in terms of $d k_1^2 = 2 k_1 dk_1$ by integrating over the azimuthal angle. This yields
\begin{align}
\frac{d \sigma^{\gamma^*_{\rm L} N \rightarrow q\bar q X}}{dz\,d k_{1}^2}\,&=\,4\,N_c \alpha_{\rm em} e_q^2\,z(1-z)\,\epsilon_f^2\,\int d^2 \mathbf{R}\,\dhd^2 \mathbf{r}_1\,J_0(k_1 r)\,\Big(K_0(\epsilon_f r_1)\,\frac{1}{k_1^2 + \epsilon_f^2}\, +\,K_0(\epsilon_f r_1)\,\frac{1}{k_1^2 + \epsilon_f^2}\, \na & - \,\,\frac{r_1}{2\epsilon_f}K_1(\epsilon_f r_1)\,\Big)N(\mathbf{r}_1,\mathbf{R})\,.
\end{align}
Assuming that the scattering amplitude has no angular dependence, $N(r_1,R)$, we can integrate over the azimuthal angle to give Eq.\,\eqref{eq:SIDIS_partonic_CGC}. A similar procedure can be repeated for the transverse polarization of the virtual photon.

\FloatBarrier
\section{Results for nuclear targets at other values of $z_h$}
\label{app:Ratio_other_zh}
We include additional results for the ratio of nuclear to proton SIDIS cross sections as a function of nuclear mass number $A$ for $z_h = 0.244$ and $z_h = 0.8$. We see that there is a larger difference between the CGC results with $x_g = x_B$ and $x_g = x_g(P_t,Q_s)$ for these values of $z_h$, compared to the previously shown $z_h=0.5$. 
This is because the $z$-integral runs from $z_h$ to 1, and the $x_g(P_t, Q_s)$ has two $z$-dependent terms that increase $x_g$ and dominate at small or large $z$, respectively. 

\begin{figure}[h]
  \begin{center}
  \includegraphics[width=0.82\linewidth]{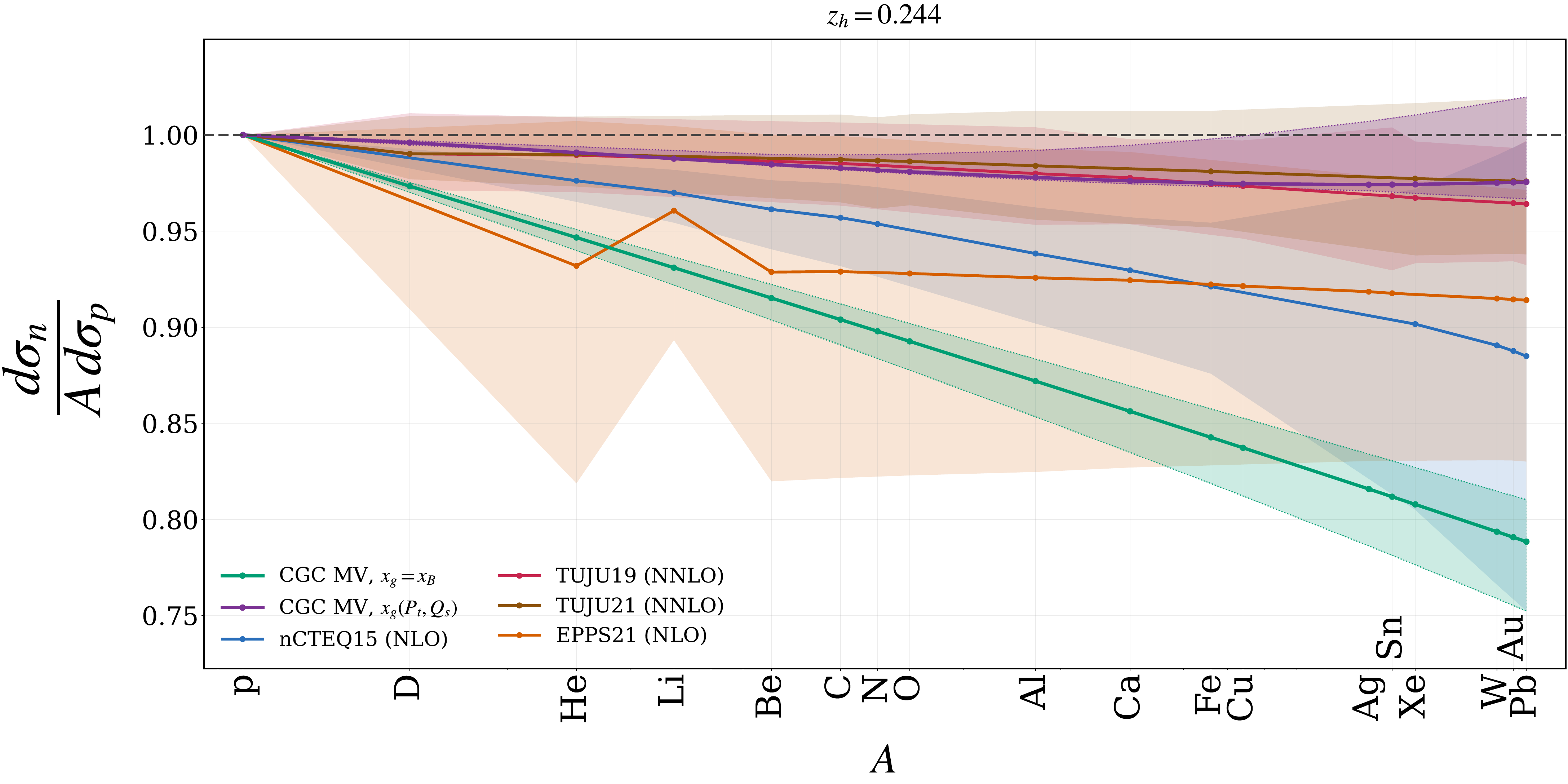}
  \end{center}
  \caption{Variation of ratio Eq.\,\eqref{eq:def_ratio_nuclear} with $A$ at $x_B = 5\times10^{-4},\,z_h=0.244, Q^2 =P_t^2= 4\,{\rm GeV}^2$ and $y= 0.69$ for the nPDF sets nCTEQ15 \cite{Kovarik:2015cma}, EPPS21 \cite{Eskola:2021nhw}, TUJU19 \cite{Walt:2019slu} and TUJU21 \cite{Helenius:2021tof}. The uncertainty bands are computed similar to Fig.\,\ref{fig:a_dependence_SIDIS}.}
  \label{fig:a_dependence_SIDIS_024}
\end{figure}

\begin{figure}[h]
  \begin{center}
  \includegraphics[width=0.82\linewidth]{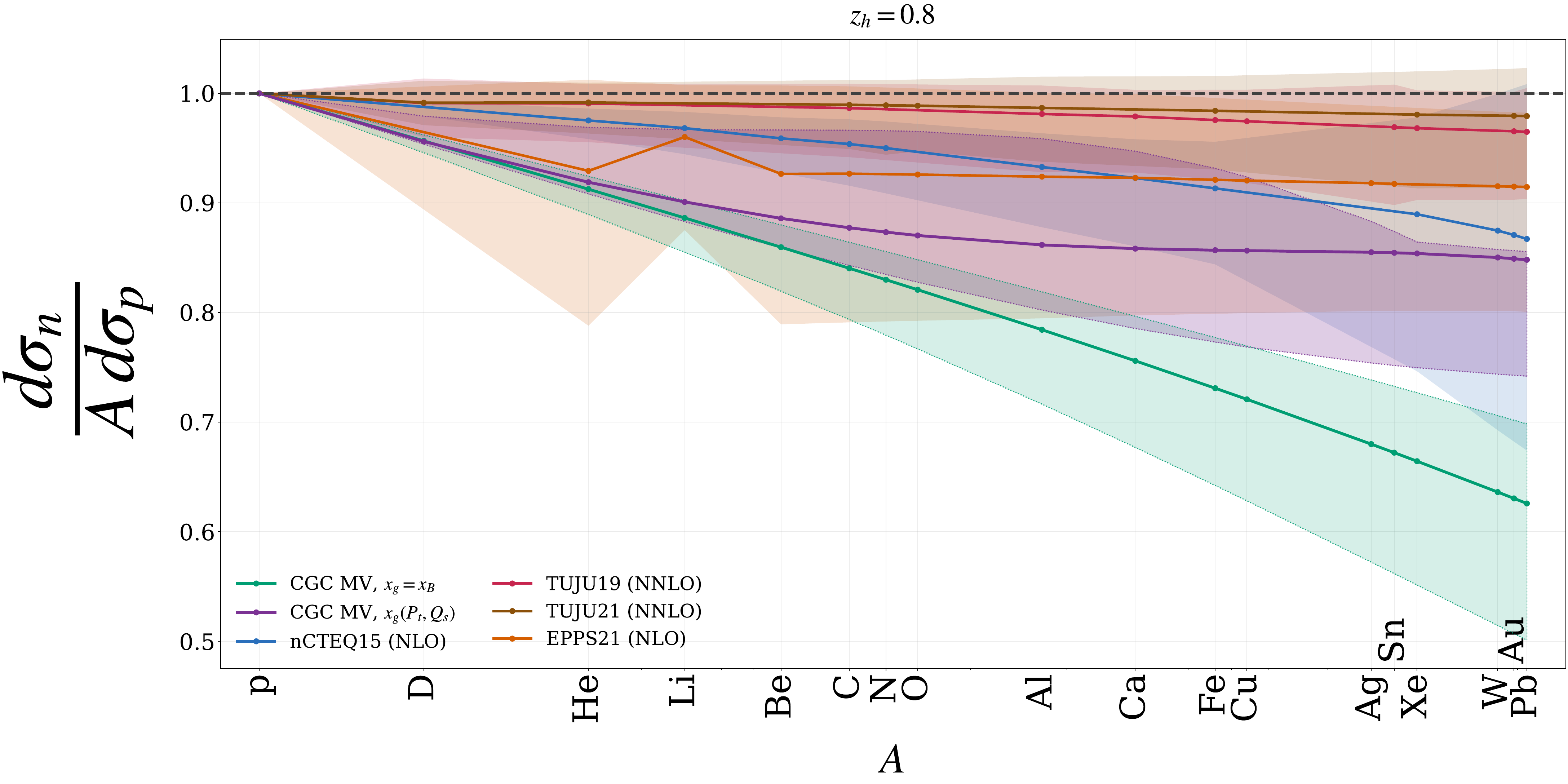}
  \end{center}
  \caption{Variation of ratio Eq.\,\eqref{eq:def_ratio_nuclear} with $A$ at $x_B = 5\times10^{-4},\,z_h=0.8, Q^2 =P_t^2= 4\,{\rm GeV}^2$ and $y= 0.69$ for the nPDF sets nCTEQ15 \cite{Kovarik:2015cma}, EPPS21 \cite{Eskola:2021nhw}, TUJU19 \cite{Walt:2019slu} and TUJU21 \cite{Helenius:2021tof}. The uncertainty bands are computed similar to Fig.\,\ref{fig:a_dependence_SIDIS}.}
  \label{fig:a_dependence_SIDIS_08}
\end{figure}

\FloatBarrier
\bibliography{SIDIS.bib}
\end{document}